\documentclass[3p, twocolumn,times,authoryear]{elsarticle}

\usepackage{hyperref}
\usepackage{lineno}
\usepackage{textcomp}
\usepackage{natbib}
\journal{Icarus}

\begin{document}

\begin{frontmatter}

\title{Wind Erosion on Mars and other Small Terrestrial Planets}

\author{Maximilian Kruss}
\ead{maximilian.kruss@uni-due.de}
\author{Grzegorz Musiolik}
\author[0000-0002-0177-7597]{Tunahan Demirci}
\author[0000-0002-7962-4961]{Gerhard Wurm}
\author{Jens Teiser}

\address{University of Duisburg-Essen, Faculty of Physics, Lotharstr. 1-21, 47057 Duisburg, Germany}

\begin{abstract}
We carried out wind tunnel experiments on parabolic flights with 100\,\textmu m Mojave Mars simulant sand. The experiments result in shear stress thresholds and erosion rates for varying g-levels at 600\,Pa pressure. Our data confirm former results on JSC Mars 1A simulant where the threshold shear stress is lower under Martian gravity than extrapolated from earlier ground-based studies which fits observations of Martian sand activity. The data are consistent with a model by Shao and Lu (2000) and can also be applied to other small terrestrial (exo)-planets with low pressure atmospheres.
\end{abstract}


\end{frontmatter}


\section{Introduction}

Dust storms are a common phenomenon on Earth. They can e.g. be observed in the Sahara, which is the largest source of aeolian soil dust and sand on Earth \citep{Schuetz1981, Goudie2001}. The causes and consequences of dust storms cannot be attributed to simple mechanisms. The behavior on the global scale is chaotic and the frequency of such storms changes over time \citep{Goudie1992}. However, conditions for dust and sand activity can be described with a microscopical force balance between gravity, cohesion, lifting wind forces and other attractive and repelling forces, e.g. a Coulomb force for single grains \citep{Shao2000, Kok2012}.

Beyond Earth, dust storms are observed on the Martian surface and have a huge impact on the planet's weather \citep{Smith2004, Heavens2011, Zurek2017}. However, aeolian particle transport on Mars is still not fully constrained as indicated by recent observations \citep{Lapotre2016, Baker2018}.
The main difficulties researching dust storm activity on Mars are the altered gravity of 0.38\,$g_{\rm{E}}$ with Earth's gravity $g_{\rm{E}} = 9.81 \,\mathrm{m}\,\mathrm{s}^{-2}$ and the low pressure of 6\,mbar. Experiments at low pressure but $1\,g_{\rm{E}}$ were carried out in the past, e.g., in the Martian surface wind tunnel at NASA Ames \citep{Greeley1976, Greeley1980, Greeley1985}.
\citet{Greeley1980} modelled the particle lifting in 0.38\,$g_{\rm{E}}$ with low-density materials. However, the inferred wind velocities needed to mobilize grains were higher than wind speeds on Mars, e.g., measured by the Viking and Phoenix landers, and could not explain particle occurrence within the Martian atmosphere \citep{Hess1977, Schofield1997, Forget1999, Holstein2010}. \citet{White1987} studied the gravity dependence on the threshold wind speed for motion initiation with a centrifuge on a parabolic flight. However, their experiments were performed under Earth's atmospheric conditions. Despite all studies in the last decades, it is still an open question which mechanisms advantage particle lifting and how strongly they contribute to it \citep{White1987, Sullivan2005, Greeley2006, Merrison2008, Sullivan2008, Kok2010a, Bridges2012}. \citet{Duran2011} and \citet{Rasmussen2015} give comprehensive overviews on aeolian sediment transport and mass fluxes. Numerical simulations often use artificially reduced thresholds for motion initiation to match the observations \citep{Daerden2015}. Effects like thermal creep due to the illumination by the sun \citep{deBeule2014}, the pressure drops within dust devils \citep{Ellehoj2010}, a lowered sediment compression in lower gravity \citep{Musiolik2018}, or different materials might support particle lifting in general.

Until today, no studies of the particle lifting mechanism for a freely chosen (low) gravity and (low) pressure exist. This is certainly attributed to the lack of small terrestrial planets with atmospheres in the Solar System in which dust and sand activity is possible. However, the number of exoplanets detected over the last years is large and the data base is continuously growing \citep{exoplanet}. Phase curves of planets or transit measurements might provide information on the surface or atmosphere of exoplanets. As planets like Mars can enshroud themselves in planet encircling dust storms, it is important to understand how particles are lifted under different gravity and different atmospheric conditions, i.e. pressure.

In this study, we significantly extend earlier work by \citet{Musiolik2018} who argue that threshold velocities needed to initiate saltation might not be as high as thought before. While that data pointed in a favorable direction, the amount of total data was rather scarce. We expand the database significantly here.  Besides measurements of threshold velocities with another sample material, we also present data in a wider gravity range of 0.15\,-\,1\,$g_{\rm{E}}$ and compare it to theoretical models. Furthermore, new aspects of erosion rates and gravity dependence are presented in this work for the first time.

\section{Experiments}

\subsection{Microgravity Setup}

\begin{figure}[h]
	\includegraphics[width=\columnwidth]{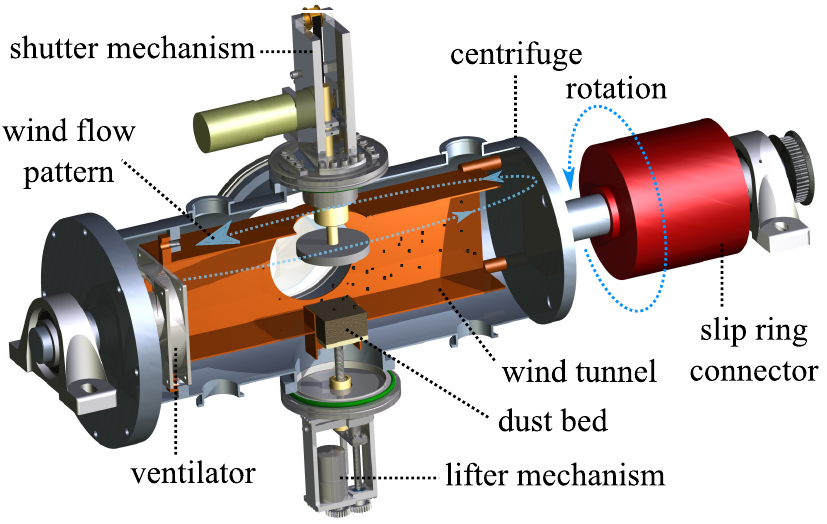}
	\caption{\label{fig.setup}Schematic of the experiment from \citet{Musiolik2018}.}
\end{figure}

The experimental setup is shown in Fig.~\ref{fig.setup} and is the same which was used by \citet{Musiolik2018} and \citet{Demirci2019}. For details, the reader is referred to these studies. However, the principle of the experiment is summarized as follows. The setup consists of a low pressure wind tunnel with a cross-section of 10~x~10\,cm which is mounted onto a centrifuge. The rotation speed determining the centrifugal forces can be set up to 2\,Hz. As the experiments were carried out under microgravity conditions during a parabolic flight with residual acceleration smaller than 0.05\,$g_{\rm{E}}$, it was possible to simulate an acceleration between around 0.15 and 1\,$g_{\rm{E}}$. The force was acting on a 4~x~4\,cm sized sample bed which was exposed to the wind flow of an adjustable fan. The maximum wind speed in the centre of the wind tunnel is around $15\,\mathrm{m}\,\mathrm{s}^{-1}$. Within the scope of this work, the wind velocity in the centre of the wind tunnel is of minor importance, as the wind velocity is traced close to the surface of the sample bed. After passing the wind tunnel, the gas can flow back through the free space between the wind tunnel and the vacuum chamber, as it is illustrated in Fig.~\ref{fig.setup}. The maximum Reynolds number inside the wind tunnel is on the order of Re\,$\approx$\,800 but the exact wind profile is not known. Since it takes some time until a certain fan velocity is established wind speeds were not varied within a parabola but kept constant.

There are two additional mechanisms to control the sample bed. A shutter, which only opens for the measurements in microgravity, covers the sample to avoid any particle spillage. The lifter unit underneath the particle bed can push the sample up to ensure that the surface of the sample is always exposed to the wind.

Prior to the parabolic flight, the chamber was filled with CO$_2$ at a pressure of around 6\,mbar which equals Martian atmospheric conditions. We note that the experiment was carried out under room temperature. Real temperatures on Mars are subject to large diurnal and seasonal variations as are pressures and it has to be kept in mind that local temperature but also pressure on Mars might differ from our experimental conditions. For each parabola a set of rotation frequency and wind speed was chosen. Potential erosion of the particle bed was observed from the side with a camera at a frame rate of 450\,fps using bright field illumination. Camera and illumination are not shown in Fig. \ref{fig.setup} for simplicity.

\subsection{Sample}

Mojave Mars Simulant (MMS), which is a common analog for Martian regolith \citep{Mojave}, was used as sample for these experiments. Obviously, choosing a realistic material for different planetary surfaces in general is not possible. However, basaltic material seems to be appropriate since it is very abundant in the Solar System and we consider it as suitable for the designed experiment.

\begin{figure}[h]
	\includegraphics[width=\columnwidth]{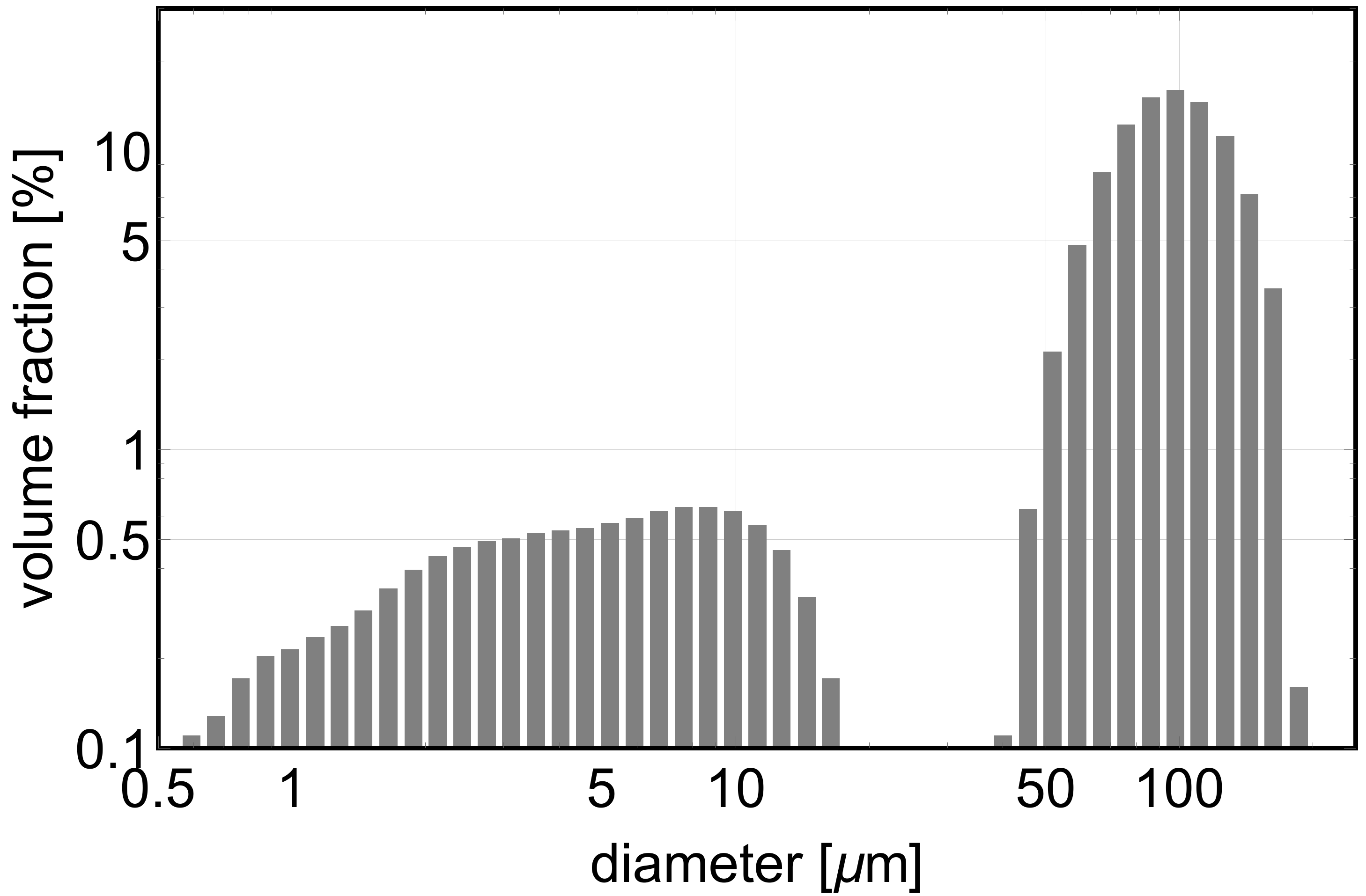}
	\caption{\label{fig.mastersizer}Grain size distribution of the used sample measured by a commercial device based on light scattering.}
\end{figure}

The sample was tempered at 200\,$^\circ\mathrm{C}$ for 24\,h to remove volatiles (especially water). It was then sieved to limit the range of grain sizes. The grain size distribution was determined using a commercial device (Mastersizer 3000, Malvern Instruments) based on light scattering. The resulting grain size distribution (see Fig.~\ref{fig.mastersizer}) is bimodal: There is a fraction of dust particles from around 1\, to \,15\,\textmu m. Noting the logarithmic scale, this is measurable but in total it is only a small contribution to the total particle volume. Besides this dust fraction there is a broad peak of sand grains around 100\,\textmu m in size. Only the sand is accessible for observations due to the limited spatial resolution of the camera. Like in \citet{Musiolik2018} the shutter lifts a large part of the sample every time it is opened. As the lifted particles fall back afterwards, the surface layers of the sample always consist of grains which settled under the respective low gravity.

\section{Results}

\subsection{Saltation Threshold}
\label{sec.threshold}

Once the wind-induced shear stress on the surface reaches a certain value, motion of individual grains is initiated. This threshold shear stress $\tau_{\rm{t}}=\rho {u_{\rm{t}}^*}^2$ is characterized by the threshold friction velocity of the wind $u^*_{\rm{t}}$ and the gas density $\rho$. By varying the fan speed in the given setup, the threshold between particle motion and no motion can be observed. It has to be noted that the wind velocity can only be changed in discrete steps.

\begin{figure}[h]
	\includegraphics[width=\columnwidth]{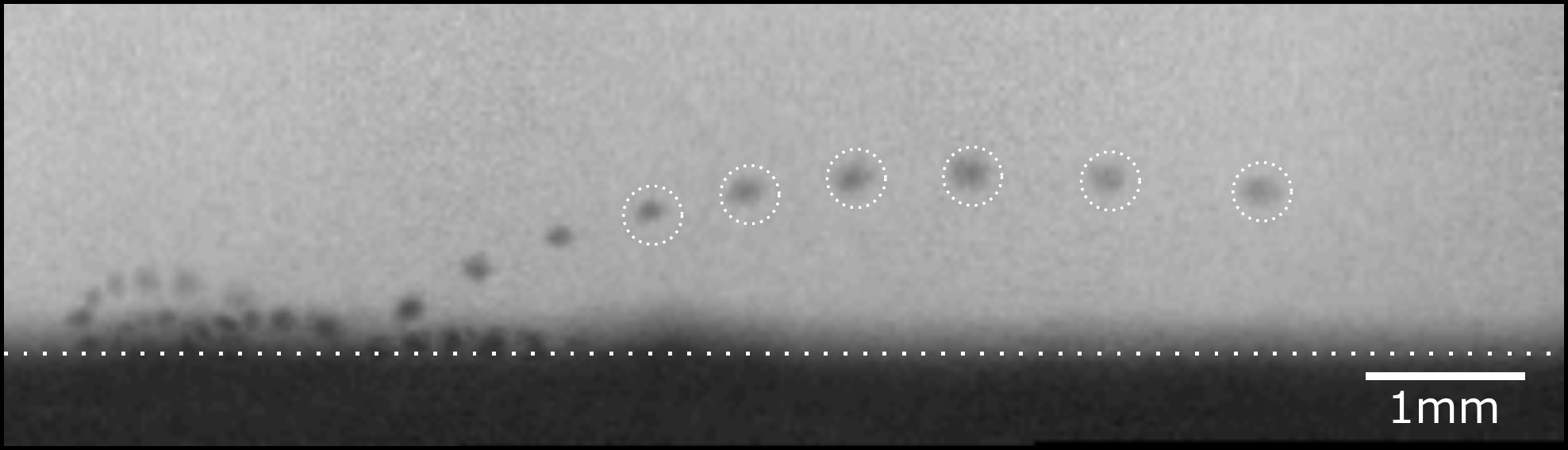}
	\caption{\label{fig.bild}Superposition of 10 subsequent frames at the threshold under Martian gravity. The encircled positions of the lifted particle were used for the fit of the gas velocity according to Eq.~\ref{eq.track}. The dotted line indicates the surface of the particle bed ($h=0$).}
\end{figure}

The applied friction velocity $u_{\rm{a}}^*$ at a given fan speed is determined by analyzing the trajectories of lifted grains. Fig.~\ref{fig.bild} shows an example trace of a lifted particle. This measurement was carried out slightly above the threshold of motion initiation. Wind speeds only lower by one discrete step did not affect the sample bed. Grains of the given size at the low pressure are not instantly coupled to the gas. They are accelerated during observation. The coupling process is characterized by a coupling time $t_{\rm{c}}$. The acceleration of the grains in flow direction ($x$-direction) at a constant height $h$ above the sample bed is, e.g., described by \citet{Wurm2001}
\begin{equation}
	\label{eq.track}
	x(t,h)=(v(h)-v_0) t_{\rm{c}}  \exp\left(-\frac{t}{t_{\rm{c}}}\right)+v(h)t+x_0,
\end{equation}
with fit parameters $v(h)$, $v_0$, $t_{\rm{c}}$ and $x_0$ being the gas velocity, the initial velocity, the coupling time and a constant, respectively. The wind profile above the surface can be mapped by tracking several grains which are lifted to different heights. We did not restrict the analysis to a certain point of time or location, but considered lifting events over several seconds and along the whole sample bed. This way any temporary fluctuations are averaged.

\begin{figure}[h]
	\includegraphics[width=\columnwidth]{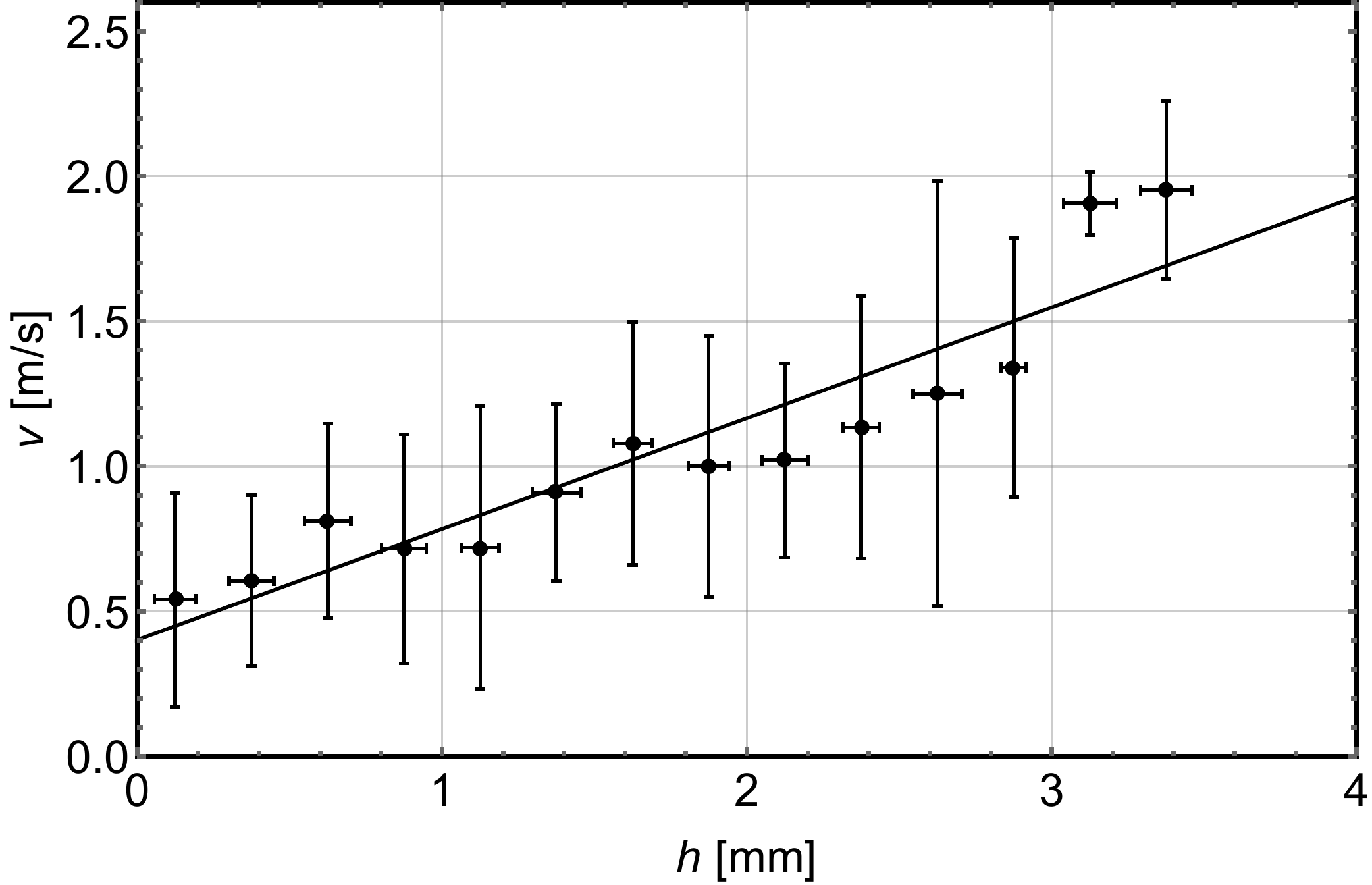}
	\caption{\label{fig.profil}Wind velocity profile at the threshold under Martian gravity. Here, a total of more than 200 individual values were binned. The error bars show the standard deviations of all values in the respective intervals. The linear fit is described by} $v(h)=(382\pm46)\,\rm{s}^{-1}\cdot \textit{h} + (0.40\pm0.09)\,\mathrm{m}\,\mathrm{s}^{-1}$.
\end{figure}

As an example, Fig.~\ref{fig.profil} shows the wind profile at the threshold under Martian gravity. The variations of the binned data reflect the uncertainty of the size of lifted grains. It was not possible to determine their exact size based on the images. The resulting data was fitted linearly because in viscous sublayers close to the ground the velocity increase can be described as linear with height \citep{Sternberg1962}. This dependence has also been applied in former work \citep{Merrison2008, Musiolik2018, Demirci2019}. The friction velocity in a Newtonian fluid is calculated according to \citep{Schlichting2006}
\begin{equation}
	u^*=\sqrt{\frac{\eta}{\rho}\frac{\textrm{d}v(h)}{\textrm{d}h}}.
\end{equation}
For CO$_2$ at room temperature and at a pressure of 6\,mbar, the dynamic viscosity and the gas density are $\eta=1.5\times 10^{-5}\,\mathrm{Pa}\,\mathrm{s}$ \citep{Laesecke2017} and $\rho=1.12\times 10^{-2}\,\mathrm{kg}\,\mathrm{m}^{-3}$, respectively. This finally yields a measured threshold friction velocity for Martian gravity of $u_{\rm{t}}^*=0.72^{+0.04}_{-0.06}\,\mathrm{m}\,\mathrm{s}^{-1}$. The uncertainty is calculated by error propagation taking into account the error of the slope in Fig.~\ref{fig.profil}. In addition, the lower limit is also affected by the stepwise increase of the wind speed. The difference of $u_{\rm{a}}^*$ between two discrete steps is around $0.02\,\mathrm{m}\,\mathrm{s}^{-1}$. If the applied friction velocity exceeds $u_{\rm{t}}^*$, the sand grains used in this experiment can be lifted under Martian conditions.

For extending the investigation from Mars to other planetary bodies and to get a better scaling on gravity in general, it is necessary to vary the gravitational forces acting on the sample bed. It was possible to determine the threshold friction velocities for five different accelerations, including a ground-based measurement. The results are summarized in Tab.~\ref{tab.results}. The uncertainty of the accelerations is caused by the residual accelerations during the parabolas.

\begin{table}[h]
	\caption{Threshold friction velocities at different gravitational accelerations.}
	\begin{tabular}{c c c}\hline
		gravity [$\mathrm{m}\,\mathrm{s}^{-2}$]&analyzed tracks&$u^*~ [\mathrm{m}\,\mathrm{s}^{-1}]$\\
		\hline 
		$1.3\pm 0.5$&127&$0.55^{+0.04}_{-0.06}$\\
		$2.5\pm 0.2$&120&$0.67^{+0.04}_{-0.06}$\\
		$3.7\pm 0.2$&208&$0.72^{+0.04}_{-0.06}$\\
		$5.7\pm 0.2$&119&$0.89^{+0.10}_{-0.12}$\\
		$9.8$&109&$0.93^{+0.06}_{-0.08}$\\
		\hline
		\label{tab.results}
	\end{tabular}
\end{table}

\begin{figure}[h]
	\includegraphics[width=\columnwidth]{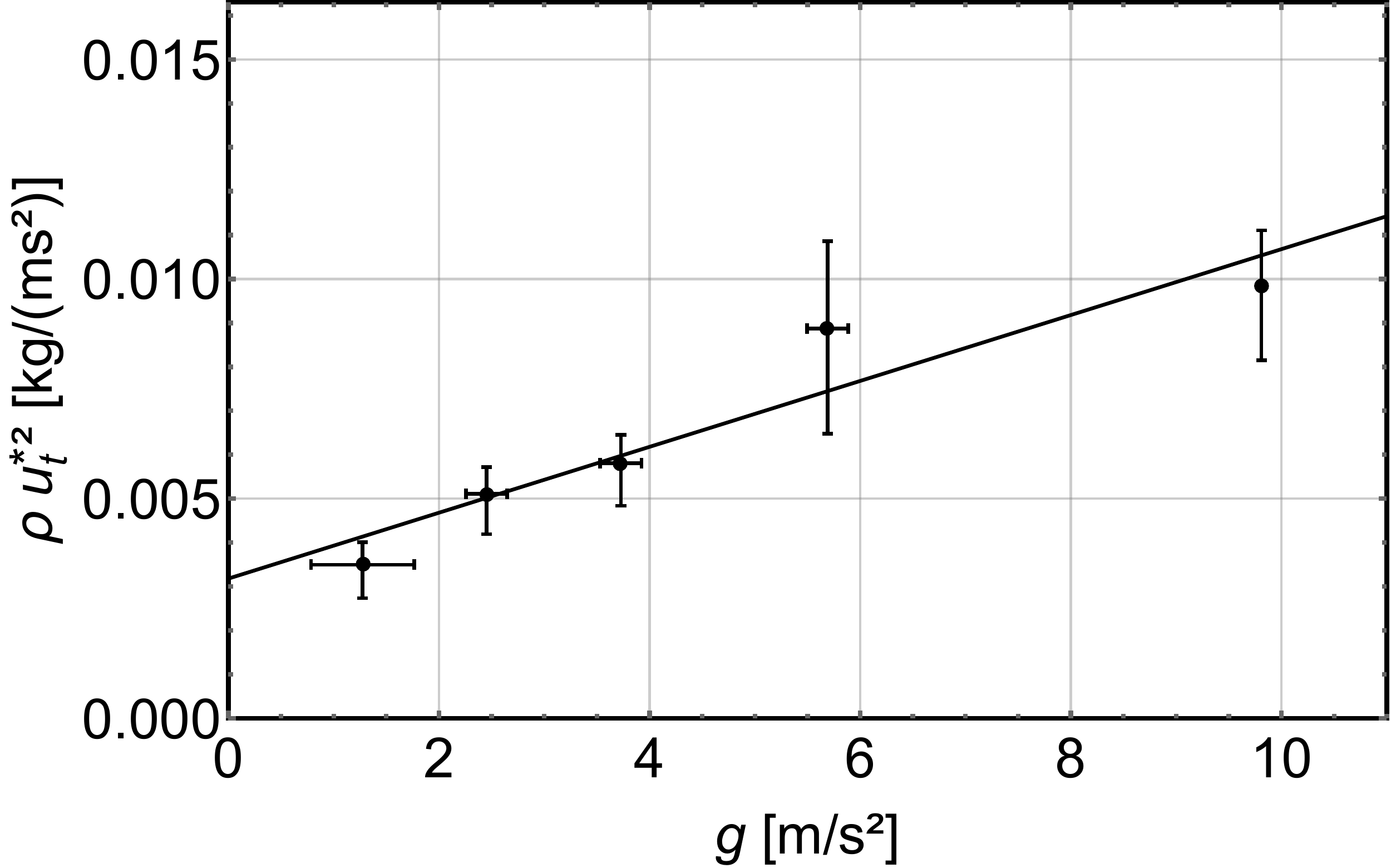}
	\caption{\label{fig.threshold}Dependence of the measured threshold shear stress on gravity $g$ overlaid by the model of \citet{Shao2000}. The linear fit is described by $\rho {u_{\rm{t}}^*}^2(g)=(7.5\pm1.5)\times 10^{-4}\,\mathrm{kg}\,\mathrm{m}^{-2}\cdot \textit{g}+(3.2\pm0.8)\times 10^{-3}\,\mathrm{kg}\,\mathrm{m}^{-1}\,\mathrm{s}^{-2}$.}
\end{figure}

\citet{Shao2000} developed an analytical expression for the dependence of $u_{\rm{t}}^*$ on gravity. In particular, they compare the shear stress necessary to lift a particle with the sum of gravity and cohesion, which has to be overcome:
\begin{equation}
	\rho {u_{\rm{t}}^*}^2=A_N\left(\rho_{\rm{p}}dg+\frac{\gamma}{d}\right).
\end{equation}
The particle's density $\rho_{\rm{p}}$, its diameter $d$ and the gravitational acceleration $g$ describe the influence of gravity, while the surface energy $\gamma$ is a measure of cohesion. $A_N$ is a dimensionless coefficient.

The linear dependence of the threshold shear stress on gravity is in agreement with the experimental data as seen in Fig.~\ref{fig.threshold}. Assuming a mean particle size of $d=100$\,\textmu m, the linear fit in Fig.~\ref{fig.threshold} yields $A_N=(2.7\pm0.5)\times10^{-3}$ and $\gamma=(1.2\pm0.5)\times10^{-4}$\,J/m$^2$. Again, the uncertainties are obtained by error propagation considering the error of the fit in Fig.~\ref{fig.threshold}. We note, that these values are calculated for monodisperse particles with $d=100$\,\textmu m. As shown in Fig.~\ref{fig.mastersizer}, this is an approximation to our sample. Nonetheless, we consider the calculated values to be useful to give an estimate of the magnitude of $A_N$ and $\gamma$. The fit also indicates that, for the given sample of irregular shaped MMS particles, cohesion plays an important role as it dominates over gravity below an acceleration of around $4\,\mathrm{m}\,\mathrm{s}^{-2}$. This behavior stands in contrast to the result of \citet{Demirci2019} who used glass beads of a much larger diameter. As expected, the reaction to wind exposure of that sample is mainly dominated by gravity.

\subsection{Erosion Rates}

\begin{figure}[h]
	\includegraphics[width=\columnwidth]{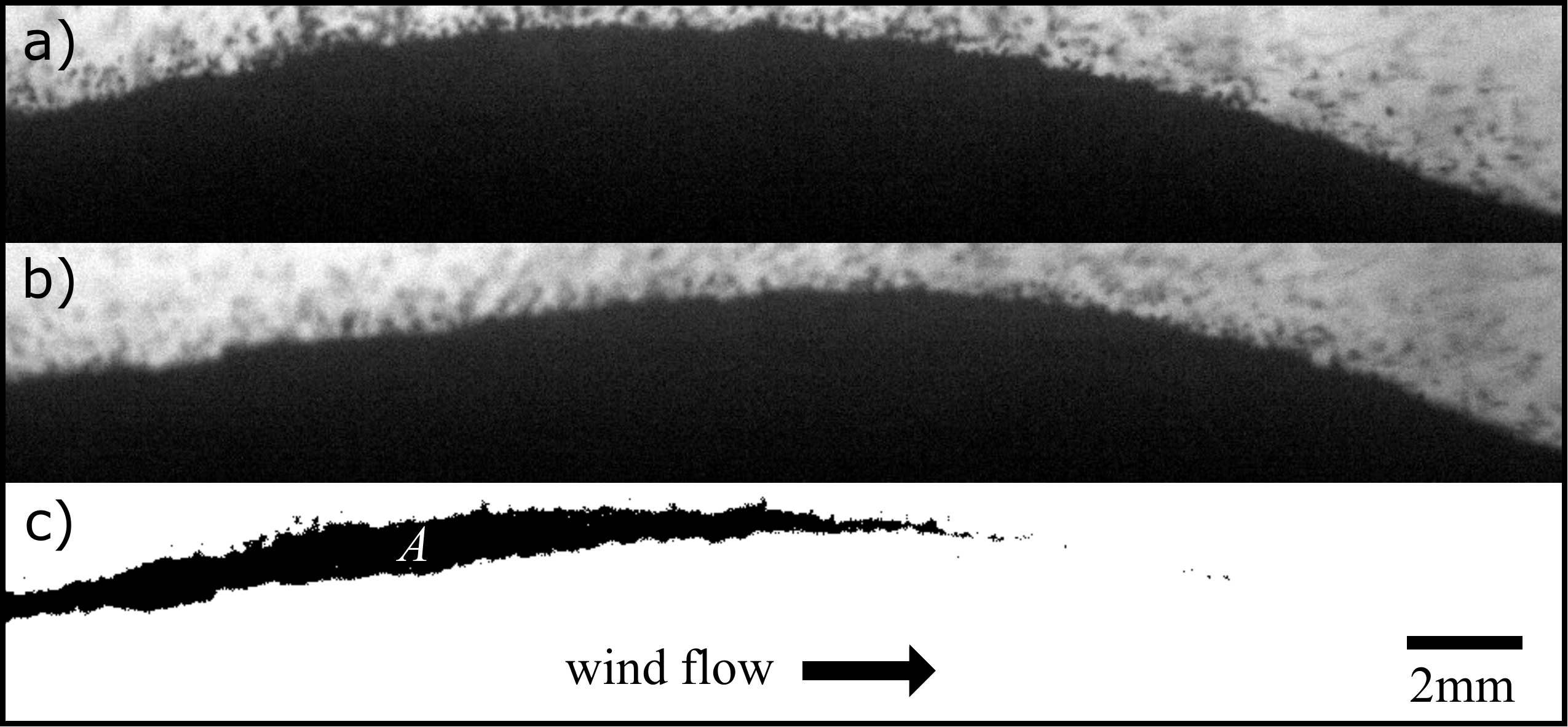}
	\caption{\label{fig.erosion}Erosion at the windward side with an applied wind velocity of $u_{\rm{a}}^*=1.03\,\mathrm{m}\,\mathrm{s}^{-1}$ and $g=2.5\,\mathrm{m}\,\mathrm{s}^{-2}$. a) and b) show the shape of the surface at the beginning of the measurement and around 11s later while c) is the difference of both images. The size of the black area in c) is defined as $A$ and displays the eroded material.}
\end{figure}

\begin{figure}[h]
	\includegraphics[width=\columnwidth]{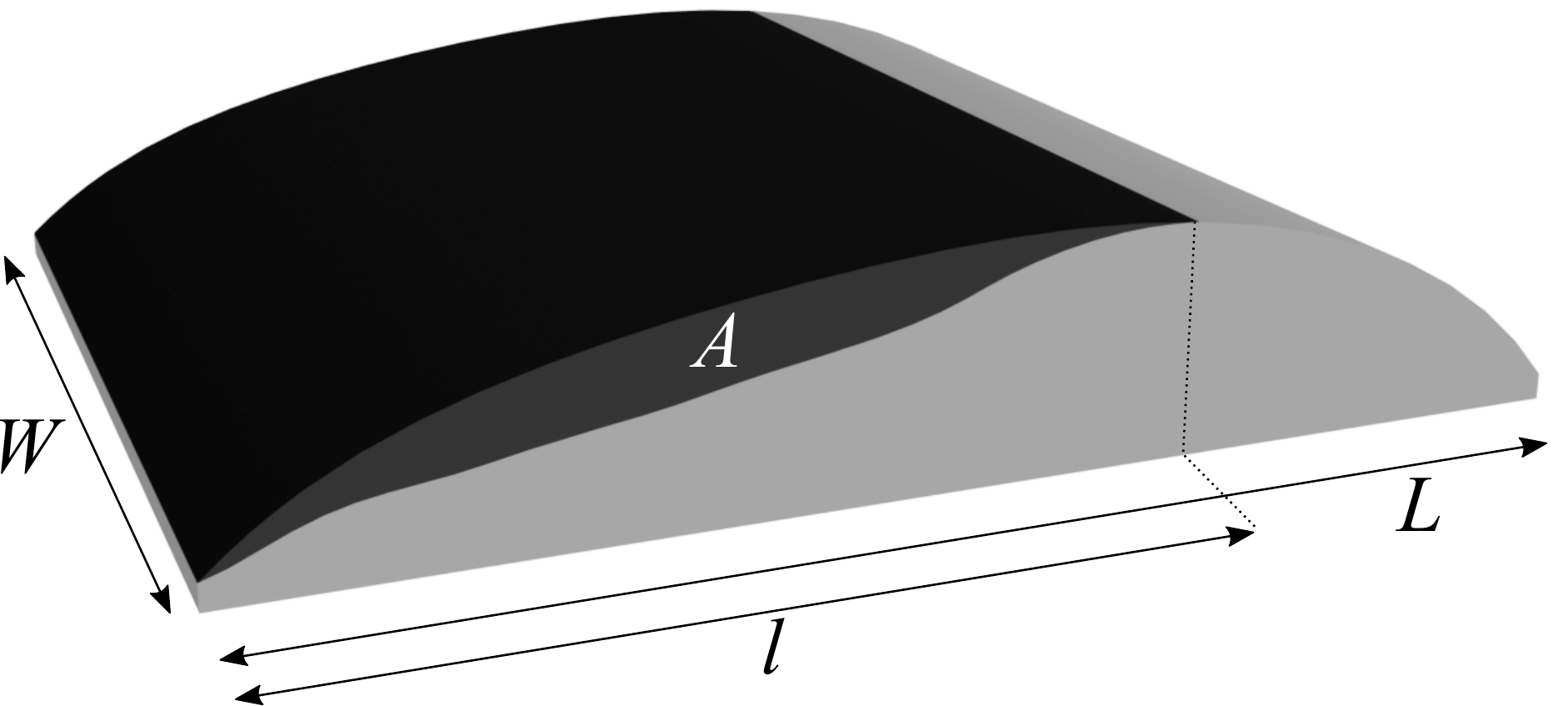}
	\caption{\label{fig.huegel}Schematic view of the assumed dimensions of the particle bed and the eroded area $A$ which can be extracted from the image data. Here, the eroded volume is marked in dark gray.}
\end{figure}

In addition to the threshold wind velocities for motion initiation, also the erosion rates for stronger winds were measured. If $u_{\rm{a}}^*>u_{\rm{t}}^*$ grains are removed and the shape of the surface changes as Fig.~\ref{fig.erosion} illustrates. Panels (a) and (b) show raw images of the samples with grains being lifted by the wind drag (panel (a) the start of the experiment and (b) the end of the experiment). To determine the eroded mass at a given time the initial image at time zero and the image at the given time are binarized and the difference image is calculated as shown in panel (c). The eroded material appears black in the difference image. The cross-sectional area $A$ of this eroded material is used to quantify the erosion rate. As only the two-dimensional cross section is visible,  the real 3D shape of the heap is not accessible for observation and its extent in the third dimension has to be estimated. Here, the particle bed is modeled as depicted in Fig.~\ref{fig.huegel}. While $W$ and $L$ describe the dimensions of the heap, $l$ measures the extent of the heap where erosion is observed. For simplicity, in this model the eroded volume is assumed to be $V=A \cdot W$. We note that this is an overestimation as the real shape is not known and the cross section along $W$ might be curved. The volume $V$ can be converted to the eroded mass using the bulk density $\rho$ of the sample which was measured on ground to $\rho = 1110\,\mathrm{kg}\,\mathrm{m}^{-3}$. The eroded mass alone is not a suitable quantity to compare different measurements. Due to the shutter of the sample bed, the heap cannot repeatedly be prepared in the same way and its size varies. To take that into account, the eroded mass is normalized to the active surface area where erosion is observed. This yields

\begin{equation}
M_{\rm{norm}} = \frac{\rho \cdot V}{l \cdot W} = \frac{\rho \cdot A}{l}.
\end{equation}

\begin{figure}[h]
	\includegraphics[width=\columnwidth]{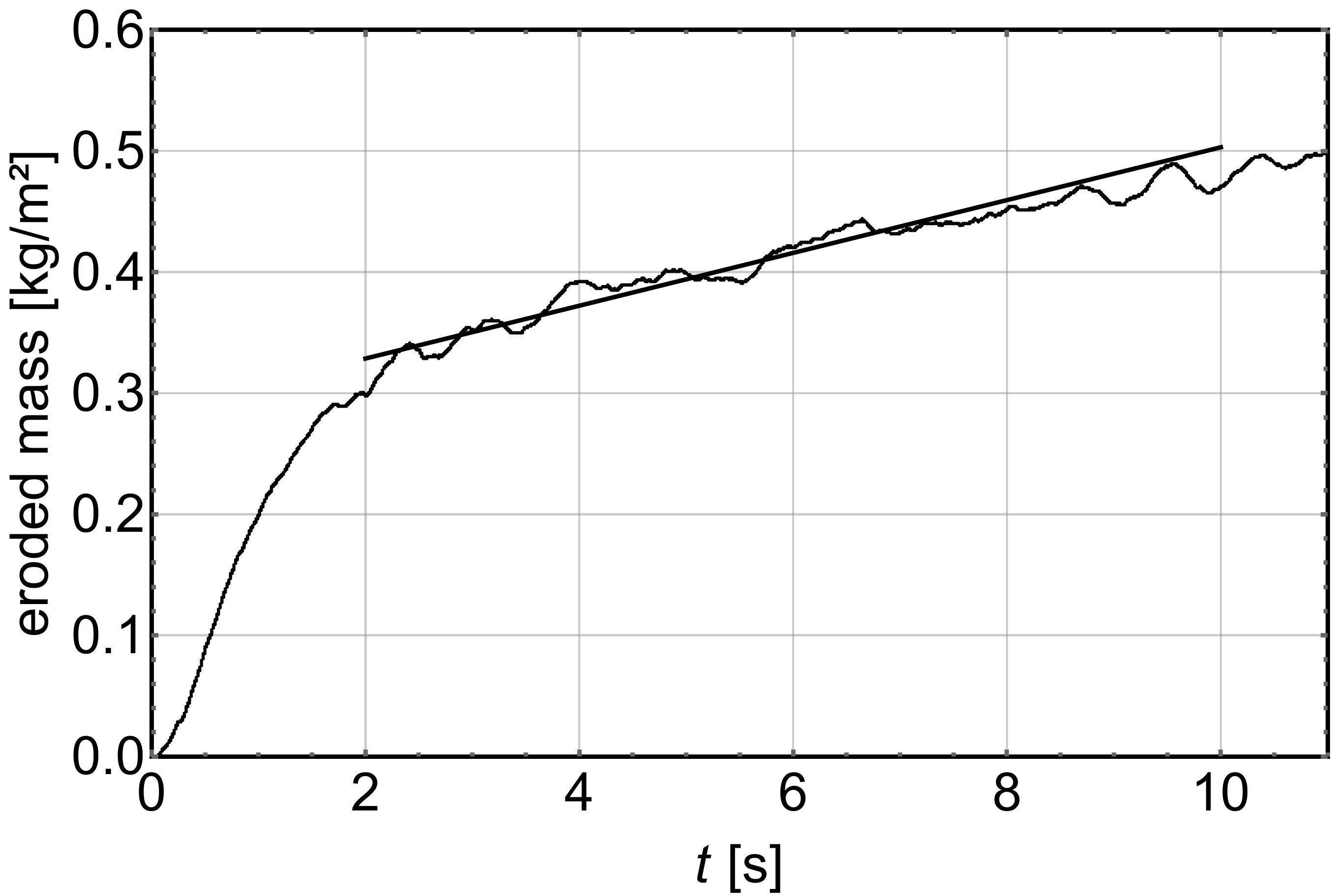}
	\caption{\label{fig.verlauf}Evolution of the cumulative eroded mass normalized to the surface area where erosion is observed. The linear fit yields an erosion rate of $\epsilon \approx 0.02\,\mathrm{kg}\,\mathrm{m}^{-2}\,\mathrm{s}^{-1}$. The applied wind velocity was $u_{\rm{a}}^*=1.03\,\mathrm{m}\,\mathrm{s}^{-1}$ at an acceleration of $g=2.5\,\mathrm{m}\,\mathrm{s}^{-2}$.}
\end{figure}

Fig.~\ref{fig.verlauf} shows the temporal evolution of the cumulated eroded mass during the measurement shown in Fig.~\ref{fig.erosion}. Once the shutter opens, a large amount of grains is entrained and settles back shortly after. During the first seconds, this loose layer is removed easily until the windward side is adapted to the wind exposure. Then, typical slope angles of the particle bed are around 6\,to 10\,$^\circ$ in our experiments. This adaptation period is mirrored by the disequilibrium in Fig.~\ref{fig.verlauf}. The number of saltating grains grows abruptly leading to a steep increase of the eroded mass until an equilibrium is established with linear mass loss. This is equivalent to a constant erosion rate $\epsilon$. Slight variations of the eroded mass over time appear due to the accuracy of the binarization (see Fig.~\ref{fig.erosion}c). This analysis is not sensitive to particle size. The erosion process is not constrained to individual grains but whole layers of material are removed.

\begin{figure}[h]
	\includegraphics[width=\columnwidth]{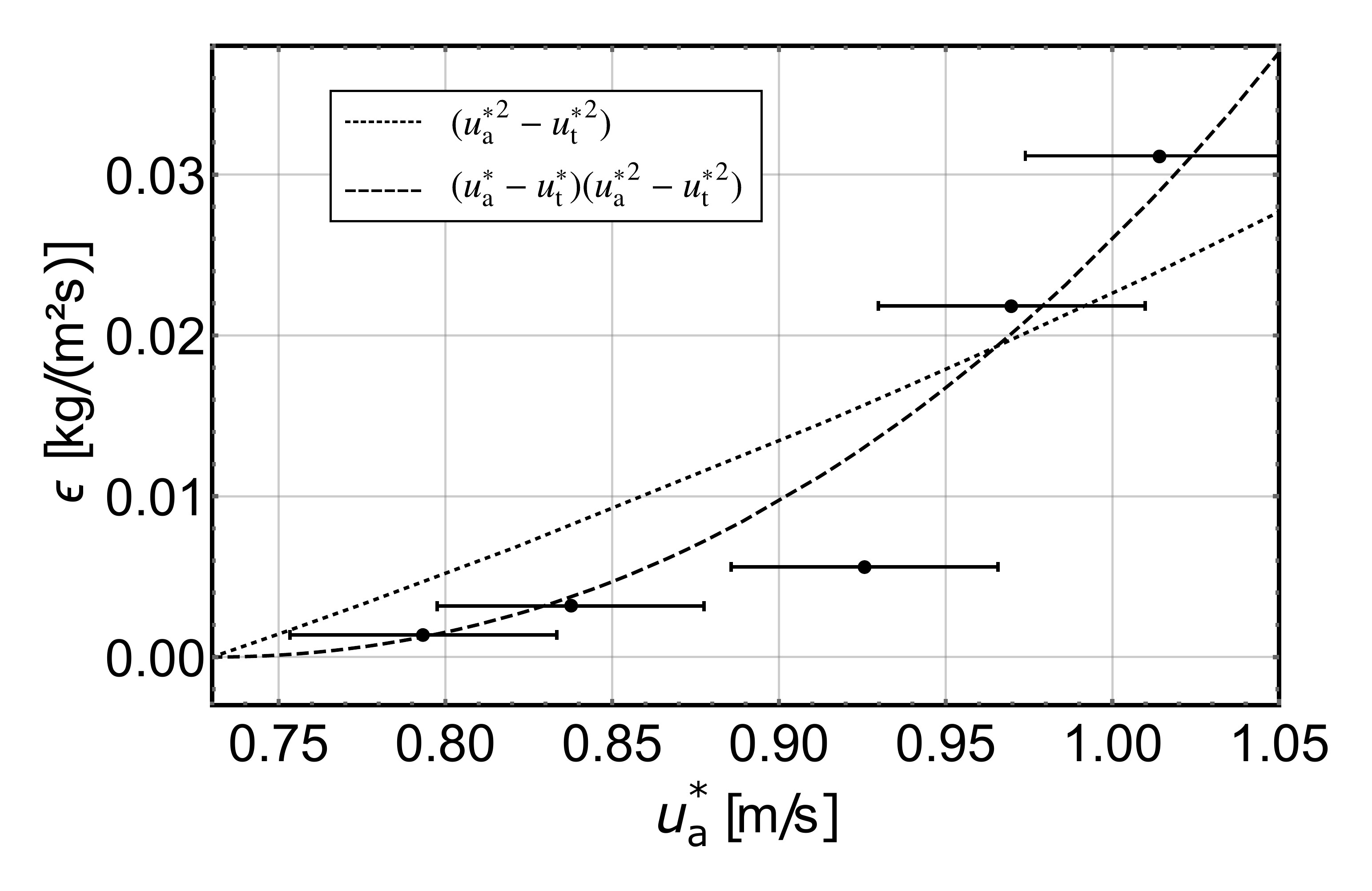}
	\caption{\label{fig.erosionmars}Erosion rates under Martian gravity ($u_{\rm{t}}^*=0.73\,\mathrm{m}\,\mathrm{s}^{-1}$) at various applied wind velocities.}
\end{figure}

At first, the erosion rate was studied under Martian conditions. The wind speeds were varied between the threshold value $u_{\rm{t}}^*=0.73\,\mathrm{m}\,\mathrm{s}^{-1}$ and the maximum speed. As expected, erosion gets stronger at larger wind speeds above the threshold, as seen in Fig.~\ref{fig.erosionmars}.

Recent models predict the erosion rate to be proportional to the excess shear stress $\epsilon \sim \tau_{\rm{a}}-\tau_{\rm{cr}}$ \citep{Martin2017}. The critical stress $\tau_{\rm{cr}}$ is mostly associated to the impact or dynamical threshold, which is lower than the fluid threshold $\tau_{\rm{t}}$ \citep{Kok2010b}. This quantity takes into account that the saltation process can be sustained by impacts of entrained particles \citep{Duran2011}. The impact treshold is not accessible in our experiment since the measurement time in one parabola is too short to change the wind speed during a parabola. Here, erosion can only be related to the fluid threshold. \citet{Creyssels2009} and \citet{Duran2011} find a quadratic dependence on the applied friction velocity in wind tunnel experiments. Fig.~\ref{fig.erosionmars} includes a fit scaling linearly with the excess shear stress and thus quadratically with the applied friction velocity. Furthermore, the data is also fitted to a cubic dependence on the velocity as was proposed by \citet{Ho2011} for a rigid particle bed. \citet{Duran2011} also find a cubic scaling for velocities well above the threshold. It has to be noted that, here, the fluid threshold is used instead of the impact threshold.\\

\begin{figure}[h]
	\includegraphics[width=\columnwidth]{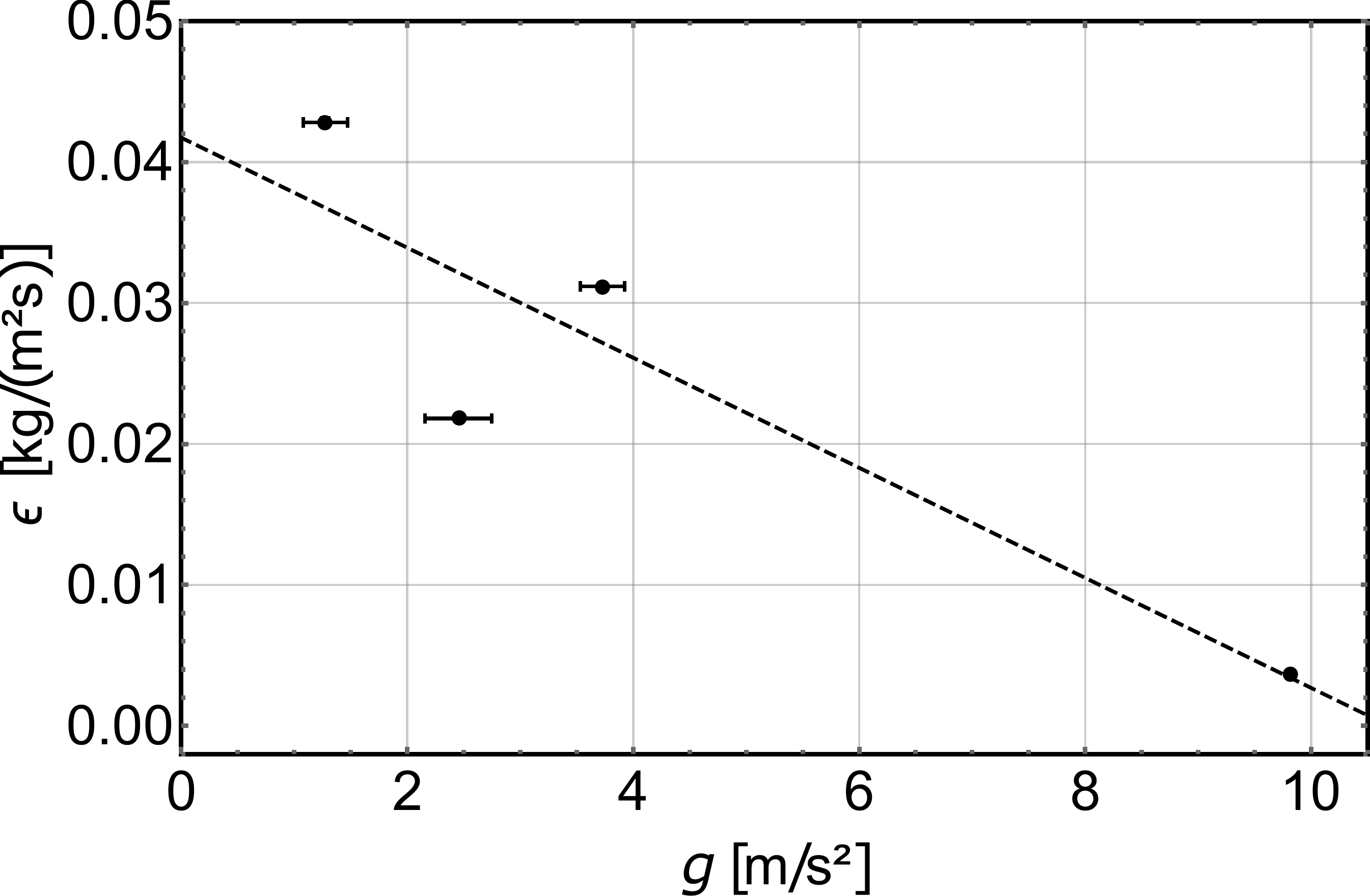}
	\caption{\label{fig.erosionsrate}Erosion rates at fixed applied wind velocity $u_{\rm{a}}^*=1.03\,\mathrm{m}\,\mathrm{s}^{-1}$ and varying gravitational acceleration. The trend line scales linearly with gravity.}
\end{figure}

By applying the analysis to measurements carried out under varying acceleration, it is possible to investigate the influence of gravity on the erosion rate. The applied wind velocity was set to the maximum value. The results are summarized in Fig.~\ref{fig.erosionsrate}. The determined erosion rates are accurate to a percent level. The large variations are due to different particle bed conditions. Nevertheless, spanning an order of magnitude, the data show a clear decrease of erosion rate with gravity as expected. A trend line scaling linearly with gravity is included.

\section{Discussion and Conclusion}

The threshold friction velocity for wind induced motion of Mojave Mars Simulant grains was measured under Martian conditions. The values are even slightly lower than those obtained by \citet{Musiolik2018} for JSC sand of a comparable size distribution. Thus, the results presented here extend the range of materials which allow motion initiation at lower wind speeds than thought before. Simulations of sand transport on Mars could therefore rely on lower thresholds without artificially reducing these values.

The general dependence between threshold shear stress and gravity below 1\,$g_{\rm{E}}$ and for mbar pressure could be studied experimentally. For the given sample and in the mbar pressure regime the data can be explained by the theoretical model of \citet{Shao2000}. There are also several bodies in the Solar System showing wind phenomena. Of course, neither the chosen gas type and pressure nor the used sample material are representative for other planetary bodies than Mars. However, the results give an estimate of how particle motion is triggered on surfaces of exoplanets with lower gravity.

To understand particle transport and to make any predictions about weather phenomena on those bodies, it is also crucial to know erosion rates. The evaluation of the amount of eroded particles presented here is far from being exact due to the lack of 3D information and varying shapes of the particle bed. However, the eroded mass can be estimated by using the model presented in Fig.~\ref{fig.huegel}. A more thorough analysis of the erosion process would require a better measurement of the eroded mass as well as a longer observation time, which is a limited resource in parabolic flights. The residual acceleration of the aircraft and the need to lift the sample during the microgravity phase usually limit the observation time to less than 10\,s.

The data are compared to two different models with quadratic and cubic dependence on $u_{\rm{a}}^*$. As the impact threshold cannot be measured here, the models differ from those by \citet{Ho2011} and \citet{Martin2017}. According to the data the erosion rate seems to grow rather cubic-like under Martian conditions. \citet{Kok2010b} found an impact threshold to maintain saltation which is an order of magnitude lower than the threshold $u_{\rm{t}}^*$ to initiate the saltation process. This large difference cannot be confirmed by this experiment as the exact value for the impact threshold was not measured. Making a more accurate statement would require more and better measurements. The characteristics of the erosion process are not representative for all local Martian conditions, but our data may give an idea of the change of erosion with speed for shallow slopes.

Finally, the presented study is a first investigation of how erosion rates are influenced by lower gravity which has not been subject of experimental work so far.

\section*{Acknowledgements}
The experiment was funded by DLR space administration under grants 50~WM~1542 and 50~WM~1760. We appreciate access to the parabolic flights which was granted by ESA. M. K. is supported by the DFG under grant WU~321/14-1.

\section*{References}

\bibliography{bib}

\begin{thebibliography}{}

\bibitem[Baker et~al., 2018]{Baker2018}
Baker, M.~M., Newman, C.~E., Lapotre, M. G.~A., Sullivan, R., Bridges, N.~T.,
  and Lewis, K.~W. (2018).
\newblock Coarse sediment transport in the modern martian environment.
\newblock {\em Journal of Geophysical Research: Planets}, 123(6):1380--1394.

\bibitem[{Bridges} et~al., 2012]{Bridges2012}
{Bridges}, N.~T., {Bourke}, M.~C., {Geissler}, P.~E., {Banks}, M.~E., {Colon},
  C., {Diniega}, S., {Golombek}, M.~P., {Hansen}, C.~J., {Mattson}, S.,
  {McEwen}, A.~S., {Mellon}, M.~T., {Stantzos}, N., and {Thomson}, B.~J.
  (2012).
\newblock {Planet-wide sand motion on Mars}.
\newblock {\em Geology}, 40:31--34.

\bibitem[Creyssels et~al., 2009]{Creyssels2009}
Creyssels, M., Dupont, P., El~Moctar, A.~O., Valance, A., Cantat, I., Jenkins,
  J.~T., Pasini, J.~M., and Rasmussen, K.~R. (2009).
\newblock Saltating particles in a turbulent boundary layer: experiment and
  theory.
\newblock {\em Journal of Fluid Mechanics}, 625:47--74.

\bibitem[Daerden et~al., 2015]{Daerden2015}
Daerden, F., Whiteway, J.~A., Neary, L., Komguem, L., Lemmon, M.~T., Heavens,
  N.~G., Cantor, B.~A., Hébrard, E., and Smith, M.~D. (2015).
\newblock A solar escalator on mars: Self-lifting of dust layers by radiative
  heating.
\newblock {\em Geophysical Research Letters}, 42(18):7319--7326.

\bibitem[{de Beule} et~al., 2014]{deBeule2014}
{de Beule}, C., {Wurm}, G., {Kelling}, T., {K{\"u}pper}, M., {Jankowski}, T.,
  and {Teiser}, J. (2014).
\newblock {The martian soil as a planetary gas pump}.
\newblock {\em Nature Physics}, 10:17--20.

\bibitem[Demirci et~al., 2019]{Demirci2019}
Demirci, T., Kruss, M., Teiser, J., Bogdan, T., Jungmann, F., Schneider, N.,
  and Wurm, G. (2019).
\newblock Are pebble pile planetesimals doomed?
\newblock {\em Monthly Notices of the Royal Astronomical Society}, page stz107.

\bibitem[Dur\'{a}n et~al., 2011]{Duran2011}
Dur\'{a}n, O., Claudin, P., and Andreotti, B. (2011).
\newblock On aeolian transport: Grain-scale interactions, dynamical mechanisms
  and scaling laws.
\newblock {\em Aeolian Research}, 3(3):243 -- 270.

\bibitem[{Ellehoj} et~al., 2010]{Ellehoj2010}
{Ellehoj}, M.~D., {Gunnlaugsson}, H.~P., {Taylor}, P.~A., {Kahanp{\"a}{\"a}},
  H., {Bean}, K.~M., {Cantor}, B.~A., {Gheynani}, B.~T., {Drube}, L., {Fisher},
  D., {Harri}, A.~M., {Holstein-Rathlou}, C., {Lemmon}, M.~T., {Madsen}, M.~B.,
  {Malin}, M.~C., {Polkko}, J., {Smith}, P.~H., {Tamppari}, L.~K., {Weng}, W.,
  and {Whiteway}, J. (2010).
\newblock {Convective vortices and dust devils at the Phoenix Mars mission
  landing site}.
\newblock {\em Journal of Geophysical Research (Planets)}, 115:E00E16.

\bibitem[exoplanet.eu, 2019]{exoplanet}
exoplanet.eu (2019).
\newblock The extrasolar planets encyclopaedia.

\bibitem[Forget et~al., 1999]{Forget1999}
Forget, F., Hourdin, F., Fournier, R., Hourdin, C., Talagrand, O., Collins, M.,
  Lewis, S.~R., Read, P.~L., and Huot, J.-P. (1999).
\newblock Improved general circulation models of the martian atmosphere from
  the surface to above 80 km.
\newblock {\em Journal of Geophysical Research: Planets},
  104(E10):24155--24175.

\bibitem[Goudie and Middleton, 2001]{Goudie2001}
Goudie, A. and Middleton, N. (2001).
\newblock Saharan dust storms: nature and consequences.
\newblock {\em Earth-Science Reviews}, 56(1):179 -- 204.

\bibitem[Goudie and Middleton, 1992]{Goudie1992}
Goudie, A.~S. and Middleton, N.~J. (1992).
\newblock The changing frequency of dust storms through time.
\newblock {\em Climatic Change}, 20(3):197--225.

\bibitem[Greeley and Iversen, 1985]{Greeley1985}
Greeley, R. and Iversen, J.~D. (1985).
\newblock Wind as a geological process on earth, mars, venus and titan.
\newblock {\em New York: Cambridge University Press}.

\bibitem[Greeley et~al., 1980]{Greeley1980}
Greeley, R., Leach, R., White, B., Iversen, J., and Pollack, J. (1980).
\newblock Threshold windspeeds for sand on mars: Wind tunnel simulations.
\newblock {\em Geophysical Research Letters}, 7(2):121--124.

\bibitem[Greeley et~al., 2006]{Greeley2006}
Greeley, R., Whelley, P.~L., Arvidson, R.~E., Cabrol, N.~A., Foley, D.~J.,
  Franklin, B.~J., Geissler, P.~G., Golombek, M.~P., Kuzmin, R.~O., Landis,
  G.~A., Lemmon, M.~T., Neakrase, L. D.~V., Squyres, S.~W., and Thompson, S.~D.
  (2006).
\newblock Active dust devils in gusev crater, mars: Observations from the mars
  exploration rover spirit.
\newblock {\em Journal of Geophysical Research: Planets}, 111(E12).

\bibitem[Greeley et~al., 1976]{Greeley1976}
Greeley, R., White, B., Leach, R., Iversen, J., and Pollack, J. (1976).
\newblock Mars: Wind friction speeds for particle movement.
\newblock {\em Geophysical Research Letters}, 3(8):417--420.

\bibitem[Heavens et~al., 2011]{Heavens2011}
Heavens, N.~G., Richardson, M.~I., Kleinböhl, A., Kass, D.~M., McCleese,
  D.~J., Abdou, W., Benson, J.~L., Schofield, J.~T., Shirley, J.~H., and
  Wolkenberg, P.~M. (2011).
\newblock Vertical distribution of dust in the martian atmosphere during
  northern spring and summer: High-altitude tropical dust maximum at northern
  summer solstice.
\newblock {\em Journal of Geophysical Research: Planets}, 116(E1).

\bibitem[Hess et~al., 1977]{Hess1977}
Hess, S.~L., Henry, R.~M., Leovy, C.~B., Ryan, J.~A., and Tillman, J.~E.
  (1977).
\newblock Meteorological results from the surface of mars: Viking 1 and 2.
\newblock {\em Journal of Geophysical Research (1896-1977)}, 82(28):4559--4574.

\bibitem[Ho et~al., 2011]{Ho2011}
Ho, T.~D., Valance, A., Dupont, P., and Ould El~Moctar, A. (2011).
\newblock Scaling laws in aeolian sand transport.
\newblock {\em Phys. Rev. Lett.}, 106:094501.

\bibitem[Holstein-Rathlou et~al., 2010]{Holstein2010}
Holstein-Rathlou, C., Gunnlaugsson, H.~P., Merrison, J.~P., Bean, K.~M.,
  Cantor, B.~A., Davis, J.~A., Davy, R., Drake, N.~B., Ellehoj, M.~D., Goetz,
  W., Hviid, S.~F., Lange, C.~F., Larsen, S.~E., Lemmon, M.~T., Madsen, M.~B.,
  Malin, M., Moores, J.~E., Nørnberg, P., Smith, P., Tamppari, L.~K., and
  Taylor, P.~A. (2010).
\newblock Winds at the phoenix landing site.
\newblock {\em Journal of Geophysical Research: Planets}, 115(E5).

\bibitem[Kok, 2010a]{Kok2010b}
Kok, J.~F. (2010a).
\newblock Difference in the wind speeds required for initiation versus
  continuation of sand transport on mars: Implications for dunes and dust
  storms.
\newblock {\em Phys. Rev. Lett.}, 104:074502.

\bibitem[Kok, 2010b]{Kok2010a}
Kok, J.~F. (2010b).
\newblock An improved parameterization of wind-blown sand flux on mars that
  includes the effect of hysteresis.
\newblock {\em Geophysical Research Letters}, 37(12).

\bibitem[Kok et~al., 2012]{Kok2012}
Kok, J.~F., Parteli, E. J.~R., Michaels, T.~I., and Karam, D.~B. (2012).
\newblock The physics of wind-blown sand and dust.
\newblock {\em Reports on Progress in Physics}, 75(10):106901.

\bibitem[Laesecke and Muzny, 2017]{Laesecke2017}
Laesecke, A. and Muzny, C.~D. (2017).
\newblock Reference correlation for the viscosity of carbon dioxide.
\newblock {\em Journal of Physical and Chemical Reference Data}, 46(1):013107.

\bibitem[Lapotre et~al., 2016]{Lapotre2016}
Lapotre, M. G.~A., Ewing, R.~C., Lamb, M.~P., Fischer, W.~W., Grotzinger,
  J.~P., Rubin, D.~M., Lewis, K.~W., Ballard, M.~J., Day, M., Gupta, S.,
  Banham, S.~G., Bridges, N.~T., Des~Marais, D.~J., Fraeman, A.~A., Grant,
  J.~A., Herkenhoff, K.~E., Ming, D.~W., Mischna, M.~A., Rice, M.~S., Sumner,
  D.~Y., Vasavada, A.~R., and Yingst, R.~A. (2016).
\newblock Large wind ripples on mars: A record of atmospheric evolution.
\newblock {\em Science}, 353(6294):55--58.

\bibitem[{Martin} and {Kok}, 2017]{Martin2017}
{Martin}, R.~L. and {Kok}, J.~F. (2017).
\newblock {Wind-invariant saltation heights imply linear scaling of aeolian
  saltation flux with shear stress}.
\newblock {\em Science Advances}, 3:e1602569.

\bibitem[Merrison et~al., 2008]{Merrison2008}
Merrison, J., Bechtold, H., Gunnlaugsson, H., Jensen, A., Kinch, K., Nornberg,
  P., and Rasmussen, K. (2008).
\newblock An environmental simulation wind tunnel for studying aeolian
  transport on mars.
\newblock {\em Planetary and Space Science}, 56(3):426 -- 437.

\bibitem[Musiolik et~al., 2018]{Musiolik2018}
Musiolik, G., Kruss, M., Demirci, T., Schrinski, B., Teiser, J., Daerden, F.,
  Smith, M.~D., Neary, L., and Wurm, G. (2018).
\newblock Saltation under martian gravity and its influence on the global dust
  distribution.
\newblock {\em Icarus}, 306:25 -- 31.

\bibitem[Peters et~al., 2008]{Mojave}
Peters, G.~H., Abbey, W., Bearman, G.~H., Mungas, G.~S., Smith, J.~A.,
  Anderson, R.~C., Douglas, S., and Beegle, L.~W. (2008).
\newblock Mojave mars simulant - characterization of a new geologic mars
  analog.
\newblock {\em Icarus}, 197(2):470 -- 479.

\bibitem[Rasmussen et~al., 2015]{Rasmussen2015}
Rasmussen, K.~R., Valance, A., and Merrison, J. (2015).
\newblock Laboratory studies of aeolian sediment transport processes on
  planetary surfaces.
\newblock {\em Geomorphology}, 244:74 -- 94.
\newblock Laboratory Experiments in Geomorphology 46th Annual Binghamton
  Geomorphology Symposium 18-20 September 2015.

\bibitem[Schlichting and Gersten, 2006]{Schlichting2006}
Schlichting, H. and Gersten, K. (2006).
\newblock {\em Grenzschicht-Theorie}.
\newblock Springer.

\bibitem[Schofield et~al., 1997]{Schofield1997}
Schofield, J.~T., Barnes, J.~R., Crisp, D., Haberle, R.~M., Larsen, S.,
  Magalh{\~a}es, J.~A., Murphy, J.~R., Seiff, A., and Wilson, G. (1997).
\newblock The mars pathfinder atmospheric structure investigation/meteorology
  (asi/met) experiment.
\newblock {\em Science}, 278(5344):1752--1758.

\bibitem[Sch\"utz et~al., 1981]{Schuetz1981}
Sch\"utz, L., Jaenicke, R., and Pietrek, H. (1981).
\newblock {Saharan dust transport over the North Atlantic Ocean}.
\newblock In {\em {Desert Dust: Origin, Characteristics, and Effect on Man}}.
  Geological Society of America.

\bibitem[Shao and Lu, 2000]{Shao2000}
Shao, Y. and Lu, H. (2000).
\newblock A simple expression for wind erosion threshold friction velocity.
\newblock {\em Journal of Geophysical Research: Atmospheres},
  105(D17):22437--22443.

\bibitem[Smith, 2004]{Smith2004}
Smith, M.~D. (2004).
\newblock Interannual variability in tes atmospheric observations of mars
  during 1999–2003.
\newblock {\em Icarus}, 167(1):148 -- 165.
\newblock Special Issue on DS1/Comet Borrelly.

\bibitem[Sternberg, 1962]{Sternberg1962}
Sternberg, J. (1962).
\newblock A theory for the viscous sublayer of a turbulent flow.
\newblock {\em Journal of Fluid Mechanics}, 13(2):241--271.

\bibitem[Sullivan et~al., 2008]{Sullivan2008}
Sullivan, R., Arvidson, R., Bell~III, J.~F., Gellert, R., Golombek, M.,
  Greeley, R., Herkenhoff, K., Johnson, J., Thompson, S., Whelley, P., and
  Wray, J. (2008).
\newblock Wind-driven particle mobility on mars: Insights from mars exploration
  rover observations at “el dorado” and surroundings at gusev crater.
\newblock {\em Journal of Geophysical Research: Planets}, 113(E6).

\bibitem[{Sullivan} et~al., 2005]{Sullivan2005}
{Sullivan}, R., {Banfield}, D., {Bell}, J.~F., {Calvin}, W., {Fike}, D.,
  {Golombek}, M., {Greeley}, R., {Grotzinger}, J., {Herkenhoff}, K.,
  {Jerolmack}, D., {Malin}, M., {Ming}, D., {Soderblom}, L.~A., {Squyres},
  S.~W., {Thompson}, S., {Watters}, W.~A., {Weitz}, C.~M., and {Yen}, A.
  (2005).
\newblock {Aeolian processes at the Mars Exploration Rover Meridiani Planum
  landing site}.
\newblock {\em Nature}, 436:58--61.

\bibitem[White et~al., 1987]{White1987}
White, B., Greeley, R., Leach, R., and Iversen, J. (1987).
\newblock Saltation threshold experiments conducted under reduced gravity
  conditions.
\newblock {\em 25th AIAA Aerospace Sciences Meeting}, page 621.

\bibitem[Wurm et~al., 2001]{Wurm2001}
Wurm, G., Blum, J., and Colwell, J.~E. (2001).
\newblock A new mechanism relevant to the formation of planetesimals in the
  solar nebula.
\newblock {\em Icarus}, 151(2):318 -- 321.

\bibitem[Zurek, 2017]{Zurek2017}
Zurek, R.~W. (2017).
\newblock {\em Understanding Mars and Its Atmosphere}, page 3â€“19.
\newblock Cambridge Planetary Science. Cambridge University Press.

\end{thebibliography}

\end{document}